\documentclass[aps,preprint]{revtex4}

\usepackage{amssymb}
\usepackage{amsmath}
\usepackage{epsfig}
\usepackage{epstopdf}
\usepackage{bm}
\usepackage{graphicx,epsfig}
\usepackage{mathrsfs}
\usepackage{dcolumn}
\usepackage{color}
\usepackage{natbib}
\usepackage{CJK}
\allowdisplaybreaks[2]

\begin{document}
\title{Topological holographic quench dynamics in a synthetic dimension}
\author{Danying Yu$^{1}$, Bo Peng$^{1}$, Xianfeng Chen$^{1,3,4,5}$, Xiong-Jun Liu$^{2,6,\dagger}$, and Luqi Yuan$^{1,*}$}
\affiliation{$^1$State Key Laboratory of Advanced Optical Communication Systems and Networks, School of Physics and Astronomy, Shanghai Jiao Tong University, Shanghai 200240, China \\
$^2$International Center for Quantum Materials and School of Physics, Peking University, Beijing 100871, China\\
$^3$Shanghai Research Center for Quantum Sciences, Shanghai 201315, China\\
$^4$Jinan Institute of Quantum Technology, Jinan 250101, China\\
$^5$Collaborative Innovation Center of Light Manipulations and Applications, Shandong Normal University, Jinan 250358, China \\
$^6$Shenzhen Institute for Quantum Science and Engineering, Southern University of Science and Technology, Shenzhen 518055, China\\
Corresponding authors: $^*$yuanluqi@sjtu.edu.cn; \
$^\dagger$xiongjunliu@pku.edu.cn}

\begin{abstract}
The notion of topological phases extended to dynamical systems
stimulates extensive studies,
of which the characterization of non-equilibrium topological invariants is a central issue and usually necessitates the information of quantum dynamics in both the time and spatial dimensions. Here we combine the recently developed concepts of the dynamical classification of topological phases and synthetic dimension, and propose to efficiently characterize photonic topological phases via holographic quench dynamics. 
A pseudo spin model is constructed with ring resonators in a
synthetic lattice formed by frequencies of light, and the quench
dynamics is induced by initializing a trivial state which evolves
under a topological Hamiltonian. Our key prediction is that the
complete topological information of the Hamiltonian is extracted
from quench dynamics solely in the time domain, manifesting
holographic features of the dynamics. In particular, two
fundamental time scales emerge in the quench dynamics, with one
mimicking the Bloch momenta of the topological band and the other
characterizing the {\em residue} time evolution of the state after
quench. For this a dynamical bulk-surface correspondence is
obtained in time dimension and characterizes the topology of the
spin model. This work also shows that the photonic synthetic
frequency dimension provides an efficient and powerful way to
explore the topological non-equilibrium dynamics.
\end{abstract}

\maketitle

\section{Introduction}

Discovery of topological quantum phases has revolutionized the
understanding of the fundamental phases of quantum matter and
ignited extensive research in condensed matter physics over the
past decades~\cite{Hasan2010,Qi2011,Yan2012,Chiu2016,Yan2017}. In
addition to the great progresses made for equilibrium phases, the
notion of topological phases has been extended to
far-from-equilibrium dynamical systems, with novel topological
physics being uncovered, such as the anomalous topological states
in Floquet
systems~\cite{Rudner2013,Hu2015,Mukherjee2017,Maczewsky2017,Wintersperger2020,VincentLiu2020,Zhang2020PRL}
and dynamical topology emerging in quantum
quenches~\cite{Caio2015,WilsonPRL2016,Hu2016,Wang2017,Heyl2018,Flaschner2018,Song2018,GongPRL2018,McGinley2019,QiuX2019,Hu2020,XiePRL2020,Yu2020,Lu2019,Slager2020}.
In particular, a universal dynamical bulk-surface correspondence
was predicted when quenching a system across topological
transition~\cite{Zhang2018,Zhang2019,Zhang2019b,Zhang2020PRL},
showing that the bulk topology of an equilibrium topological phase 
has a one-to-one correspondence to quench-induced dynamical
topological patterns emerging on the lower-dimensional momentum
subspaces called band inversion surfaces (BISs). The dynamical
bulk-surface correspondence connects the equilibrium topological
phases with far-from-equilibrium quantum dynamics, which was
further extended to correlated system~\cite{Zhang2019c},
high-order regimes~\cite{XLYu2020,Gong2020}, and to generic slow
non-adiabatic quenches~\cite{LiPRA2020}. This opens the way to
characterize equilibrium topological phases by non-equilibrium
quench dynamics, and inversely, to classify non-equilibrium
quantum dynamics by topological theory, with the experimental
studies having been widely reported
recently~\cite{Sun2018b,Yi2019,Wang2019,Song2019,Ji2020,Xin2020,Niu2020,BChen2021}.
The non-equilibrium topological invariants are typically defined
via time dimension and momentum space, and their characterization
naturally necessitates the information of quantum dynamics in both
the time and spatial dimensions.

As an extension of the spatial degree of freedom, the synthetic dimensions \cite{r1,r2,r3} was proposed and opened an intriguing avenue towards quantum simulation of exotic topological physics beyond physical dimensions~\cite{r4,r5,YuanNanophotonics}. 
Following the numerous theoretical proposals on synthetic
dimensions using different degree of freedoms such as the
frequency or the orbital angular momentum of
light~\cite{r6,r7,r8,r9}, and the hyperfine levels of
atoms~\cite{syn-atom-1}, experiments have been recently performed
to demonstrate the two-dimensional topological insulator
\cite{r10} and the Hall ladder \cite{r11} in the synthetic space,
where the effective magnetic field for photons is generated, and
visualize the edge states~\cite{syn-atom-2,syn-atom-3}. Further,
the high-dimensional physics can be studied in a photonic platform
with lower dimensionality \cite{r8,r12,r13,r14,r15,r16}. More
recently, the experimental platforms for generating the synthetic
dimension along the frequency axis have also been proposed and
demonstrated using the ring resonator~\cite{r7,r8,r11,r17,Li2020},
in which the photonic modes at equally-spanned frequencies are coupled through the dynamic modulation. In this system, the band structure in the synthetic dimension can be measured in the static steady-state regime in the experiment~\cite{r17}. On the other hand, with the synthetic dimensions the novel optical phenomena and applications have been be proposed, 
including the realizations of unidirectional frequency
translation~\cite{s1}, pulse narrowing~\cite{s2}, active photon
storage~\cite{s3}, and topological laser~\cite{s4}.

In this work, we combine the concepts of dynamical classification and the synthetic dimension, and propose a highly efficient scheme to characterize topological phases by holographic quench dynamics. 
We construct a one-dimensional pseudo spin model in a photonic synthetic lattice formed by frequencies of light, and investigate the quench dynamics by initializing a trivial phase which evolves under a topological Hamiltonian. 
We show that the full dynamical evolution is featured by two
fundamental time scales, with which the quench dynamics exhibit
universal topological patterns. In particular, one time scale
mimics the Bloch momenta of the topological band and the other
characterizes the {\em residue} time evolution of the state after
quench. The dynamical topological patterns obtained on BISs render
an emergent dynamical bulk-surface correspondence and provide a
holographic characterization of the topological spin model, with
the complete information being captured in the single-variable,
i.e. the time evolution. The emergent dynamical topology is robust
against disorders and has high feasibility in the implementation.
This work shows advantages in exploring the topological phases
with holographic quench dynamics in the synthetic dimensions, and
provides the insight into classifying the far-from-equilibrium
dynamics with nontrivial topology based on the synthetic photonic
crystals.

\section{Model}
\begin{figure}[htbp]
\centering
\includegraphics[width=0.95\textwidth ]{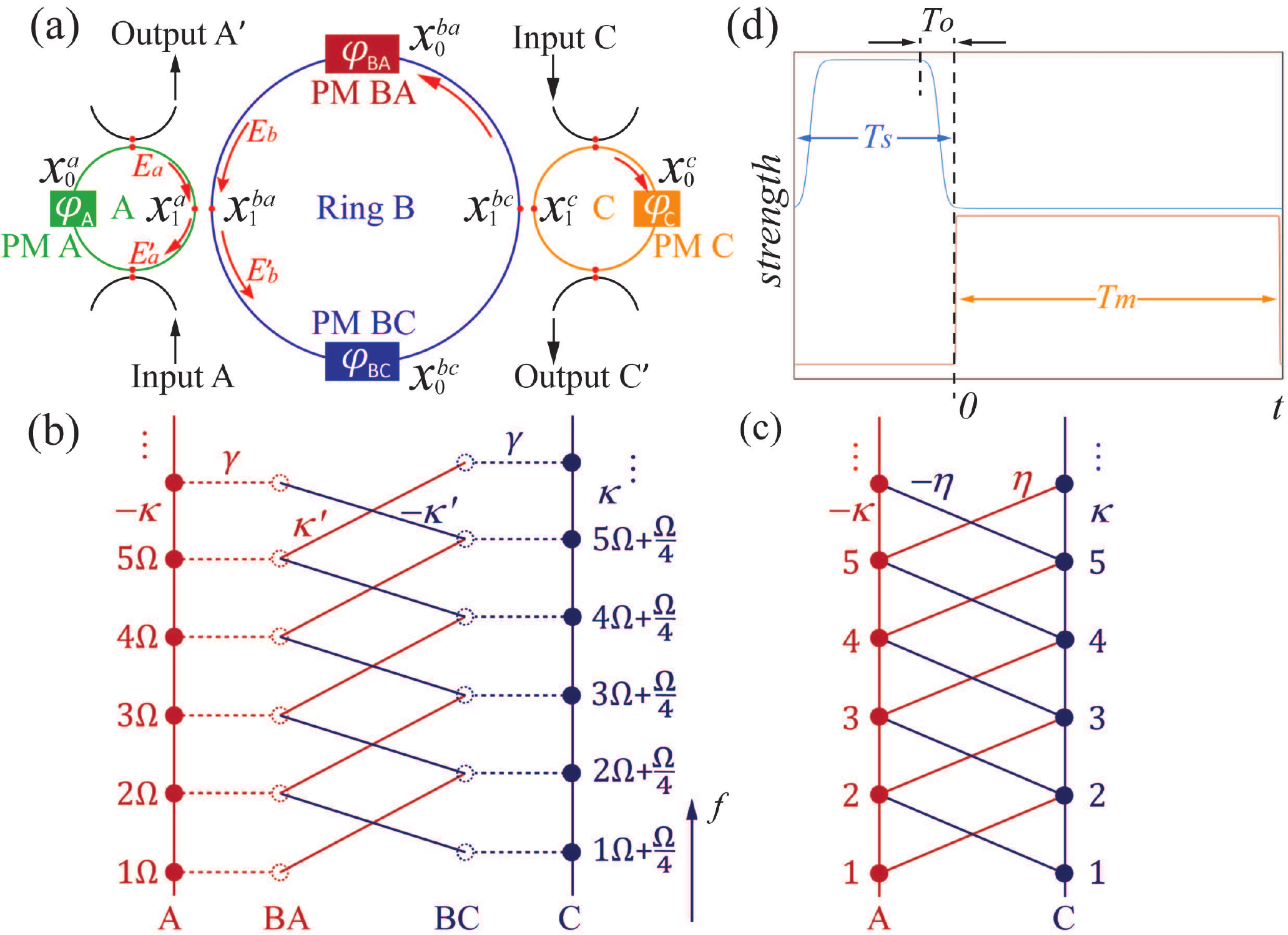}
\caption{(a) A schematic design of the ring resonator system with
phase modulators. External waveguides are used to input (output)
signal. (b) The diagram that shows couplings between modes A$_m$
(Red solid dot) and BA$_m$ (Red dashed dot) at frequencies
$\omega_{A,m}=m\Omega$ and modes C$_m$ (Blue solid dot) and BC$_m$
(Blue dashed dot) at frequencies $\omega_{C,m}=m\Omega +
\Omega/4$. Modulators induce nearest-neighbor couplings  (Solid
line) between nearby modes along the frequency axis of light,
modes in different rings are coupled through the evanescent wave
(Dashed line). $\phi_A=\phi_{BC}=\pi$ and $\phi_C=\phi_{BA}=0$ in
modulators give negative (Red line) and positive (Blue line)
couplings, respectively. (c) The effective tight-binding model of
the  pseudo-spin lattice in the nontrivial case ($\phi=\pi$). (d)
Time sequences of input source ($T_S$) and modulation ($T_m$) in
simulations. $T_O$ is the turn-on/-off time.}\label{figure.1}
\end{figure}
We start with illustrating our idea of using ring resonators under
dynamic modulations to artificially engineer a tight-binding
lattice of pseudo spin states along the frequency axis of light.
As shown in Fig.~\ref{figure.1}(a), the system under the study in
this work contains three ring resonators, with each hosting a set
of resonant frequency modes. Two of the resonators (A and C) will
be used to mimic a pseudospin-$1/2$ system. Let the group velocity
be zero in the waveguide that constructs the ring. We set that the
ring A supports a set of resonant modes at frequencies
$\omega_{A,m}=m\Omega$, where $m$ is an integer and $\Omega=2\pi
c/Ln_g$ is the free-spectral-range of the ring. Here $c$ is the
speed of light, $L$ is the circumference of the ring A, and $n_g$
is the effective refractive index. The resonant modes in the ring
C with the same circumference $L$ have frequencies:
$\omega_{C,m}=m\Omega+\Omega/4$. We use a ring B with the
circumference $4L$ to serve as an auxiliary ring \cite{r28}. The
ring B have shifted resonant modes at frequencies $\omega_{B,m} =
m\Omega/4+ \Omega/8$. Hence, though there are evanescent couplings
between nearby rings, the field resonantly circling inside the
ring A(C) is not resonant in the auxiliary ring B.

The couplings between the ring resonators are engineered by
properly setting the phase modulators. We place one phase
modulator [labelled as PM A(C) in Fig.~\ref{figure.1}(a)] inside
the ring A(C). The light that transmits through the modulator in
the ring A(C) undergoes dynamical modulation with the transmission
coefficient as \cite{a2}:
\begin{eqnarray}\label{new1}
T_{A(C)}=e^{i2\kappa \cos[\Omega_0 t+\phi_{A(C)}]},
\end{eqnarray}
where $\kappa$ is the modulation strength, $\Omega_0$ is the
modulation frequency, and $\phi_{A(C)}$ is the modulation phase in
the modulator PM A(C). We consider the resonant modulation, i.e.,
$\Omega_0=\Omega$, so each modulator couples the nearest-neighbor
resonant modes in two rings in the first-order approximation. The
ring B contains two phase modulators, which are labelled as PM BA
and PM BC with the corresponding transmission coefficients $T_1$
and $T_2$:
\begin{eqnarray}\label{new3}
T_{1(2)}=e^{i2\kappa'\cos(\Omega_{1(2)} t+\phi_{BA(BC)})},
\end{eqnarray}
where $\kappa'$ is the modulation strength, $\Omega_{1,2}$ are the
modulation frequencies, and $\phi_{BA}$ and $\phi_{BC}$ are
modulation phases in PM BA and PM BC, respectively. We set
$\Omega_1=5\Omega/4$ so that the field component at the frequency
$\omega_{A,m}$ couples with the component at $\omega_{C,m+1}$.
Similarly, for $\Omega_2=3\Omega/4$ the component at
$\omega_{A,m+1}$ couples with the component at $\omega_{C,m}$.

The pseudospin-$1/2$ system is realized by modulating the
resonator couplings. The field in the ring A(C) is coupled with
the field in the ring B through the evanescent wave. The
corresponding coupling equation is described by the coupling
matrix between input electric field amplitudes $E_a$, $E_b$ and
output amplitudes $E_a'$ and $E_b'$ labelled in
Fig.~\ref{figure.1}(a):
\begin{eqnarray}\label{2}
 \begin{pmatrix}
 E_a'\\
 E_b'
 \end{pmatrix}
 =
 \begin{pmatrix}
 \sqrt{1-\gamma^2} & -i\gamma \\
 -i\gamma & \sqrt{1-\gamma^2}
 \end{pmatrix}
 \begin{pmatrix}
E_a\\
 E_b
 \end{pmatrix}.
\end{eqnarray}
Here $\gamma$ is the coupling strength. The coupling matrix
between B and C rings follows the same expression. With the above
ingredients we can map the setting to the diagram described in
Fig. 1(b) which gives our lattice model as shown below. Here, the
description of external waveguides used for the input source and
output detections in simulations are not included.


The coupling between resonant modes in rings A and C are mediated
by the ring B, as illustrated in Fig.~\ref{figure.1}(b). The
physics is described below. The energies of the resonant modes
A$_m$ leak into the temporary non-resonant component (labelled as
BA$_m$) in the ring B, which may decay quickly. However, the
modulations characterized in Eq.~(\ref{new3}) in the ring B
convert the energies in these components BA$_m$ to other
non-resonant components (labelled as BC$_m$), and the latter
components are transferred to resonant modes C$_m$ in the ring C.
Hence the couplings to the auxiliary ring B serve as an
intermediate process which mediates an second-order coupling
between resonant modes A$_m$ and C$_m$. This process mimics the
second-order Raman process between two states through virtual
transitions to an intermediate state in quantum mechanics. On the
other hand, the resonant modes with frequencies $\omega_{A,m}$
(labelled as A$_m$) in the ring A couples between each other
through the dynamic modulation characterized in Eq.~(\ref{new1}),
forming a synthetic lattice for A itself in the frequency
dimension, similar for resonant modes C$_m$) in the ring C.

We now turn to the effective model of the coupled ring system. The modulations including modulation phases inside rings have high tunability~\cite{r28}. 
We choose modulation phases to be either $0$ or $\pi$. For
example, we can set $\phi_A=\phi_{BC}=\pi$, which gives the
corresponding negative coupling, or $\phi_C=\phi_{BA}=0$, which
gives the positive corresponding coupling. The system can be
described by an effective Hamiltonian:
\begin{eqnarray}\label{1}
 H& =&\sum\limits_m {{\omega_{A,m}}a_m^\dag {a_m} + {\omega _{C,m}}c_m^\dag {c_m}  + 2\kappa [\cos(\Omega t + \phi )(a_m^\dag {a_{m + 1}} + a_{m + 1}^\dag {a_m})}  \nonumber \\
  &&+ \cos (\Omega t)(c_m^\dag {c_{m + 1}} + c_{m + 1}^\dag {c_m})]+2\eta[\cos (5\Omega t/4)(a_m^\dagger {c_{m+1}}+c_{m+1}^\dagger {a_m}) \nonumber \\
  &&+\cos (3\Omega t/4+\phi)(c_m^\dagger {a_{m+1}}+a_{m+1}^\dagger c_m)],
\end{eqnarray}
where $a$ ($a^\dagger$) and $c$ ($c^\dagger$) are the annihilation
(creation) operators for resonant modes A$_m$ and C$_m$ in rings A
and C, respectively, $\eta=\kappa'\gamma^2$ for the weakly
coupling case, and $\phi$ can be either $\pi$ or $0$, depending on
what model we are going to study. For the case of $\phi=\pi$, it
corresponds to the diagram shown in Fig.~\ref{figure.1}(c). The
Hamiltonian can be rewritten under the rotating-wave
approximation:
\begin{eqnarray}\label{r}
H_r&=& \sum\limits_m [e^{i\phi}\kappa(a_m^\dagger a_{m+1}+a_{m+1}^\dagger a_m)+ \kappa(c_m^\dagger c_{m+1}+c_{m+1}^\dagger c_m)\nonumber\\
&&+\eta(a_m^\dagger c_{m+1}+c_{m+1}^\dagger
a_m)+e^{i\phi}\eta(c_m^\dagger a_{m+1}+a_{m+1}^\dagger c_m)].
\end{eqnarray}
Eq.~(\ref{r}) with $\phi=\pi$ describes a topological Hamiltonian
of a one-dimensional pseudospin-$1/2$ lattice model (with the
modes A and C denoting the spin-up and spin-down, respectively)
along the synthetic frequency dimension as shown in
Fig.~\ref{figure.1}(c) \cite{Zhang2019b}. In the following we
proceed to study the quench dynamics, and shall show how the
dynamical topological patterns emerge in a nontrivial way from
simulations.

\section{Simulation}
We perform simulations using the realistic model based on the
setting in Fig.~\ref{figure.1}(a). The simulation has been used to
successfully describe the dynamics of the ring-based system in the
synthetic space and is discussed in details in Refs
\cite{r7,s1,r28}. Here we briefly summarize the procedure. The
electric field inside the waveguide is \cite{b1}
\begin{eqnarray}\label{prop}
E(t,r_\perp,x)=\sum_{m} \mathcal{E}(t,x)E_m(r_\perp)e^{i\omega'_m
t},
\end{eqnarray}
where $x$ is the propagation direction along the waveguide that
composes the ring resonator, $r_\perp$ is the perpendicular
directions of $x$, $\omega'_m$ is either $\omega_{A,m}$ or
$\omega_{C,m}$, $E_m(r_\perp)$ is the modal profile for the ring A
or C as well as the auxillary ring B, and $\mathcal{E}(t,x)$ is
the associated modal amplitude in different rings. Under the
slowly varying envelope approximation, Eq.~(\ref{prop}) satisfies
the wave equation:
 \begin{eqnarray}\label{prop2}
 [\frac{\partial}{\partial x}+i\beta (\omega'_m)]\mathcal{E}_m-\frac{n_g}{c} \frac{\partial}{\partial t}\mathcal{E}_m=0,
 \end{eqnarray}
 where $\beta$ is the wavevector. The ring has the periodic boundary condition $\mathcal{E}_m(t,x+L)=\mathcal{E}_m(t,x)$ for rings A and C, and $\mathcal{E}_m(t,x+4L)=\mathcal{E}_m(t,x)$ for the ring B.

 When the light passes through the phase modulation, the field undergoes dynamic modulation and modal amplitudes obey \cite{b2}:
 \begin{eqnarray}\label{PMC}
 \mathcal{E}_m^{A/C}(t^+,x_0^{a/c})= J_0(\kappa)\mathcal{E}_m^{A/C}(t^-,x_0^{a/c})+J_1(\kappa)\mathcal{E}_{m-1}^{A/C}(t^-,x_0^{a/c})e^{i\phi_{A/C}}-J_1(\kappa)\mathcal{E}_{m+1}^{A/C}(t^-,x_0^{a/c})e^{-i\phi_{A/C}},
 \end{eqnarray}
 where $t^{\pm}=t+0^{\pm}$, $x_0^{a/c}$ represents the position of the modulator in the ring A or C in Fig.~\ref{figure.1}(a), and $J_0$ and $J_1$ are the $0$th and $1$st order Bessel functions, respectively. Here we take the first-order approximation and only consider the nearest-neighbor couplings, which turns out to be fine in this model and also in other works for weak modulations \cite{r7,s1,r28}. Similarly, dynamic modulations on both PM BA and PM BC at positions $x_0^{ba}$ and $x_0^{bc}$, respectively, are described by following equations:
 \begin{eqnarray}\label{PMBA}
  \mathcal{E}_m^{BA(BC)}(t^+,x_0^{ba(bc)})= J_0(\kappa')\mathcal{E}_m^{BA(BC)}(t^-,x_0^{ba(bc)})-J_1(\kappa')\mathcal{E}_{m+1}^{BC(BA)}(t^-,x_0^{ba(bc)})e^{-i\phi_{BA(BC)}},
 \end{eqnarray}
 Eqs.~(\ref{PMC}) and (\ref{PMBA}) reflect transmission coefficients in Eqs.~(\ref{new1}) and~(\ref{new3}), respectively. The coupling in Eq.~(\ref{2}) between fields in rings A and B through the evanescent wave at corresponding positions in Fig.~\ref{figure.1}(a) can be described by:
 \begin{eqnarray}\label{coucb1}
 \mathcal{E}_m^{A(BA)}(t^+,x_1^{a(ba)})=\sqrt{1-\gamma^2}\mathcal{E}_m^{A(BA)}(t^-,x_1^{a(ba)})-i\gamma\mathcal{E}_m^{BA(A)}(t^-,x_1^{a(ba)}),
  \end{eqnarray}
 The coupling between fields in ring C and ring B is similarly described.

 In simulations, the four external waveguides coupling the rings A and C, as shown in Fig.~\ref{figure.1}(a), are applied to input the source fields (which can also be decomposed to the frequency component $E_m^{A,\mathrm{in}}$ and $E_m^{C,\mathrm{in}}$) and detect the output signal ($E_m^{A,\mathrm{out}}$ and $E_m^{C,\mathrm{out}}$). The input/output coupling between the waveguide and the ring is also described by the similar equations~(\ref{coucb1}) with the coupling strength $\gamma'$ \cite{r28}.
 \begin{figure}[htbp]
 \centering
 \includegraphics[width=0.9\textwidth ]{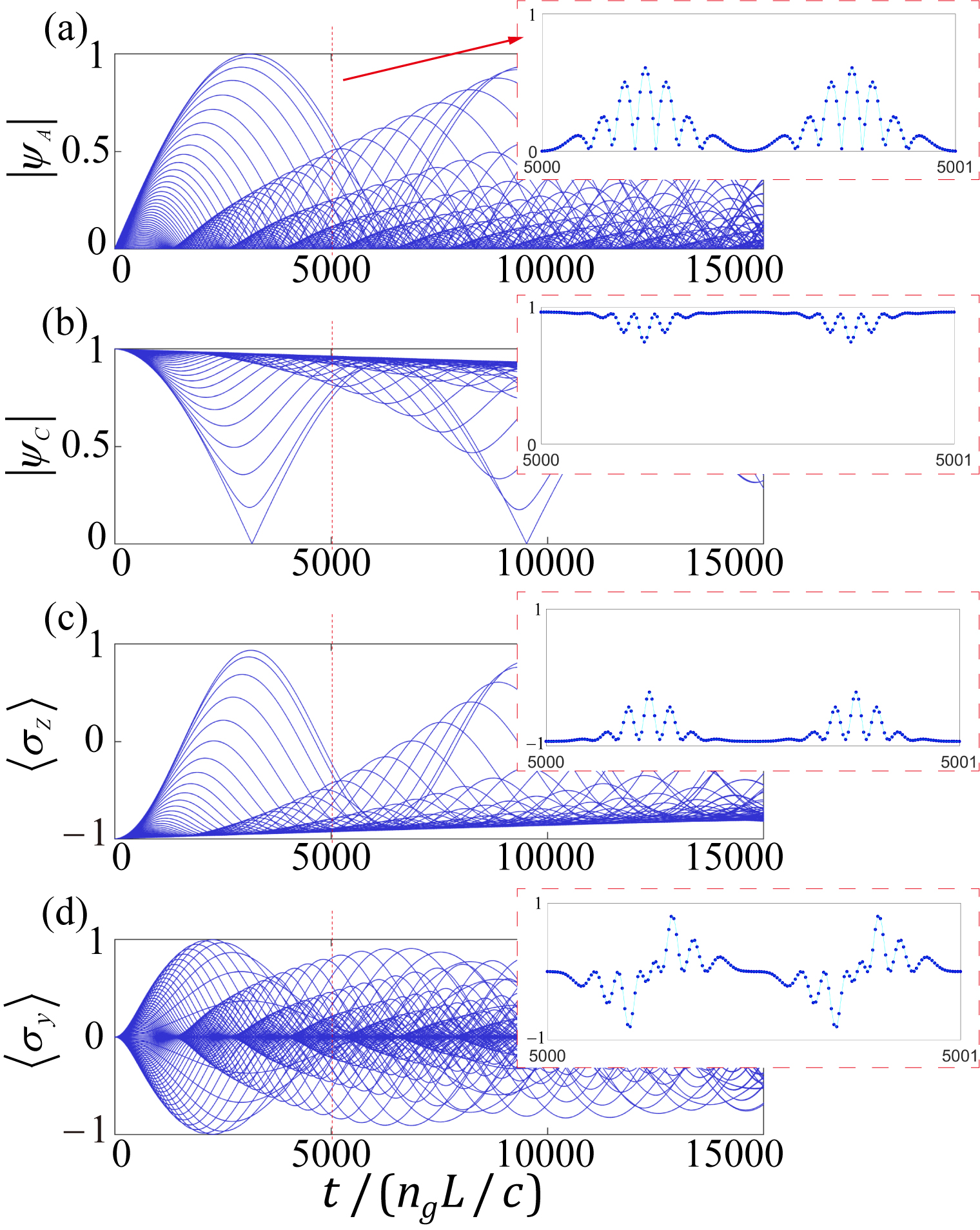}
 \caption{The electric field amplitudes and spin texture from simulations. (a) and (b) The normalized electric field amplitudes $|\psi_A|$ and $|\psi_C|$ versus discrete time collected from the output waveguides. 160 data points are collected in one roundtrip (=$1\ n_gL/c$). (c) and (d) The time evolution of normalized spin textures ($\langle \sigma_z \rangle$ and $\langle \sigma_y \rangle$). Insets are the corresponding zoom-in plots, where data points are connected with lines, showing that $|\psi_A|$, $|\psi_C|$, $\langle\sigma_z\rangle$, and $\langle\sigma_y\rangle$ are evolving continuously along the time dimension.}\label{figure.2}
 \end{figure}

 \section{Results and Analysis}
\subsection{Simulation results}
In this section, we show the feasibility of directly measuring the
bulk topology of the system quench dynamics process. To this
purpose, we first prepare the initial state of the system by
injecting a monochromic light at the center frequency
$\omega_{C,0}$ into the input external waveguide C. This source
field has the temporal form with a normalized field amplitude $s$
 \begin{eqnarray}
 E_0^{C,\mathrm{in}}=s\{\mathrm{tanh}[0.05(t+t_S-t_O/2)]+\mathrm{tanh}[0.05(-t_O/2-t)]\},
 \end{eqnarray}
 where $t_O$ is the turn-on/-off time and $T_S$ is the pulse temporal duration. This choice of the input source only excites the mode C$_{m=0}$ in the ring C. No mode in the ring A is prepared at $t=0$. Thus the initial excitation of the ring system is fully polarized, giving an initial deep trivial state \cite{Zhang2018}. The modulations are then turned on at $t=0$ in the time sequence diagram shown in Fig.~\ref{figure.1}(d) with the modulation time $T_m$. Signals from output external waveguides are collected for further analysis in our simulations. The turning-on of the modulations makes the system be characterized by the non-trivial pseudospin-$1/2$ lattice model described in Fig.~\ref{figure.1}(c), and the quench dynamics is induced with the initial state evolving under the topological Hamiltanian of Eq.~(\ref{r}).

For the simulation, we set that both ring A and ring C contain 81
resonant modes ($m=-40,-39,...,40$). The parameters designed in
the ring resonator system in Fig.~\ref{figure.1}(a) are:
$\kappa=0.0025\ c/Ln_g$, $\kappa'=0.2\ c/Ln_g$, $\gamma=0.1\
c/Ln_g$, $\gamma'=0.003\ c/Ln_g$, respectively. We also choose
$T_S=1000\ n_gL/c$, $T_O=200\ n_gL/c$, and $T_m=15000\ n_gL/c$.

Signals are collected from $t=0$ to $t=T_m$ for all the frequency
components $E_m^{A,\mathrm{out}}(t)$ and $E_m^{C,\mathrm{out}}(t)$
at both output waveguides. Therefore, the total electric field
amplitudes of the signals, $\psi_A(t)$ and $\psi_C(t)$, can be
retrieved by
$\psi_A(t)=\sum_{m}E_m^{A,\mathrm{out}}(t)e^{-i\omega_{A,m} t}$,
and $\psi_C(t)=\sum_{m}E_m^{C,\mathrm{out}}(t)e^{-i\omega_{C,m}
t}$. We plot normalized $|\psi_A(t)|$ and $|\psi_C(t)|$ under the
time evolution in Figs.~\ref{figure.2}(a) and~\ref{figure.2}(b),
respectively, which show nearly periodic patterns over the short
time [see the zoom-in plots in both figures]. Nevertheless, the
dynamics does not show the periodic stability over a long time,
which is an evidence for the steady-state solution of the system
\cite{r17}. With the collected signal, one can further construct
the spin textures $\langle
\sigma_z(t)\rangle=|\psi_A|^2-|\psi_C|^2$ and $\langle
\sigma_y(t)\rangle= -i\psi_{A}^*\cdot \psi_{C}e^{i\Omega
t/4}+i\psi_{C}^*e^{-i\Omega t/4} \cdot \psi_{A}$, with which we
shall show the essential prediction of this work that the two time
fundamental scales emerge in the dynamics and the novel
topological patterns are resulted   (see also Appendix A for
details). Time evolution of normalized $\langle
\sigma_z(t)\rangle$ and $\langle \sigma_y(t) \rangle$ are plotted
in Figs.~\ref{figure.2}(c) and~\ref{figure.2}(d). Note that the
raw pseudopsin dynamics characterized by $\langle \sigma_z\rangle$
and $\langle \sigma_y \rangle$ do not exhibit topological feature
explicitly, but actually contain the complete information as
presented below.

\subsection{Topological quench dynamics in the synthetic frequency dimension}

 \begin{figure}[htbp]
 \centering
 \includegraphics[width=0.98\textwidth ]{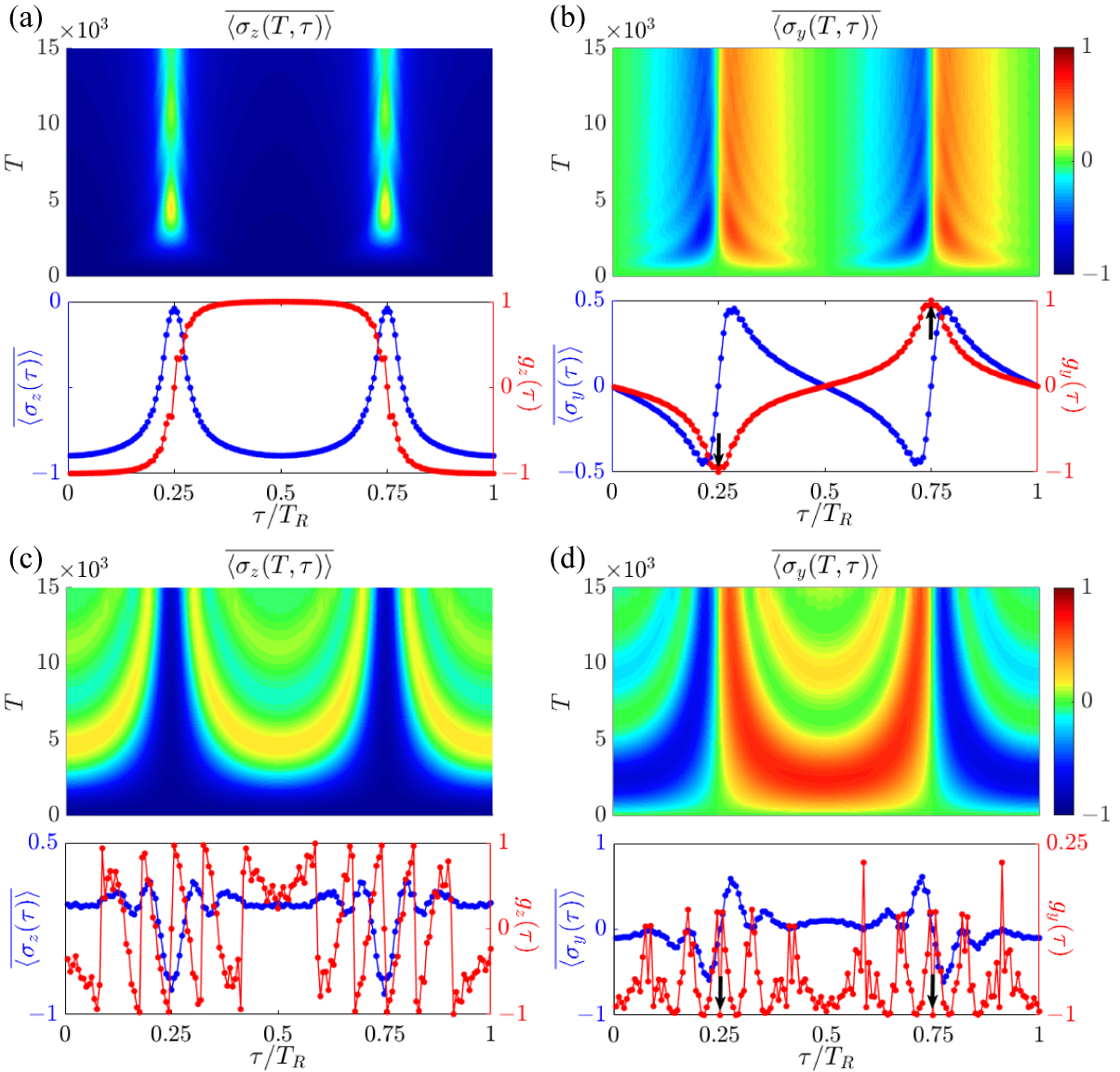}
 \caption{The time evolution of the normalized spin textures reconstructed by using two time scales ($\tau$ and $T$). (a) and (b) The evolution of averaged spin-polarization $\overline{\langle \sigma_{z,y}(T,\tau) \rangle}$, the overall spin-polarization $\overline{\langle \sigma_{z,y}(\tau)\rangle}$, the dynamical spin texture $g_{z,y}(\tau)$, respectively, with $\phi=\pi$. (c) and (d) The evolution of averaged spin-polarization $\overline{\langle \sigma_{z,y}(T,\tau) \rangle}$, the overall spin-polarization $\overline{\langle \sigma_{z,y}(\tau)\rangle}$, the dynamical spin texture $g_{z,y}(\tau)$, respectively, with $\phi=0$. Red arrows point to values of $g_y(\tau_{1,2})$.}\label{figure.3}
\end{figure}

A novel observation is that two fundamental time scales emerge in
the time evolution of the pseudospin polarization, denoted as the
slow time variable $T$ and the fast time variable $\tau$,
respectively. The real time reads $t=TT_R+\tau$, with
$T_R=2\pi/\Omega$. Thus $T$ is the round-trip numbers, which is a
discrete non-negative integer ($T=0,1,2,\ldots$), and $\tau \in
[0,T_R)$ is the round-trip time. Note that for the synthetic
dimension along the frequency axis of light, the round-trip time
$\tau$ corresponds to the Bloch momentum $k_f$, i.e. the wave
vector reciprocal to the frequency. For studies modelling a static
system, the transmission of light versus $\tau$ at the periodicity
$T_R$ can give the steady-state bandstructure of the synthetic
lattice along the frequency axis of light \cite{r11,r17}. However,
in our present study, the topological quench dynamics is extracted
from the two emergent time scales in the time dimension, of which
$\tau$ mimics the {\em Bloch momentum} and $T$ denotes the {\em
residue} time evolution of the state.

We therefore represent results $\langle \sigma_z(t)\rangle$ and
$\langle \sigma_y(t)\rangle$ by defining $\langle
\sigma_z(T,\tau)\rangle$ and $\langle \sigma_y(T,\tau)\rangle$,
which give the quench dynamics at the Bloch momenta $k_f=\tau$
evolving over the discrete residue time $T$. The dynamical
classification theory~\cite{Zhang2018} states that the quench
dynamics exhibit nontrivial topology captured by the time-averaged
spin texture in momentum space. Unlike the previous theory, here
we define the averaged spin-polarization $\overline{\langle
\sigma_z(T,\tau) \rangle}$ and $\overline{\langle
\sigma_y(T,\tau)\rangle}$ over the residue time $T$, given by
\begin{eqnarray}
 \overline{\langle \sigma_{y,z}(T,\tau) \rangle}=\frac{1}{T+1}\sum_{T'=0}^T\langle \sigma_{y,z}(T',\tau)\rangle.
\end{eqnarray}
We show $\overline{\langle \sigma_{y,z}(T,\tau) \rangle}$ and the
overall spin-polarization $\overline{\langle
\sigma_{y,z}(\tau)\rangle}\equiv \lim_{T\rightarrow\infty}
\overline{\langle \sigma_{y,z}(T,\tau) \rangle}$ in
Figs.~\ref{figure.3}(a)-\ref{figure.3}(c). The plots exhibit
nontrivial dynamical pattern characterized via the two time scales
$\tau$ and $T$. First, the overall averaged polarizations vanish
$\overline{\langle \sigma_{y,z}(\tau)\rangle}=0$ at two special
points with $\tau_1=0.25 T_R$ and $\tau_2=0.75 T_R$. Such two
characteristic points are known as band inversion points in the 1D
Brillouin zone (BZ)~\cite{Zhang2018}. Secondly, we define a new
dynamical spin texture in the following way:
\begin{eqnarray}
\vec g(\tau) =\left\{
\begin{array}{lr}
   (1/{\cal N})\partial_\tau \overline{\langle \vec\sigma(\tau)\rangle},\ \ \  \tau=\tau_{1,2}; \\
   (s/{\cal N})\overline{\langle \vec\sigma(\tau)\rangle},\ \ \ \ \ \ \mathrm{other}\ \tau\  \mathrm{points},
   \end{array}
   \right.
\end{eqnarray}
where $\cal N$ is the normalization factor, the derivative
direction is chosen from the area in-between the two band
inversion points to that out of them if $\tau$ is at band
inversion points, and $s=-1$ ($+1$) if $\tau$ is in the region
in-between (out of) the two band inversion points for other $\tau$
points. One finds that at the two band inversion points
$g_z(\tau_{1,2})=0$, while $g_y(\tau_1)=-g_y(\tau_2)=-1$ points in
opposite directions [see Fig.~\ref{figure.3}(d)], giving a nonzero
dynamical topological number, i.e. the zeroth Chern number
$C_0=[g_y(\tau_2)-g_y(\tau_1)]/2$, as defined via the two band
inversion points. This manifests the emergent dynamical
bulk-surface correspondence~\cite{Zhang2018}, and the bulk
topology of the Hamiltonian of the synthetic lattice constructed
in Fig.~\ref{figure.1}(c) is topologically nontrivial. As a
comparison, we can use the same system but change the modulation
phase $\phi=0$, and show the numerical results in
Figs.~\ref{figure.3}(e)-\ref{figure.3}(h). The emergent dynamical
field $\vec g(\tau)$ at two band inversion points are the same
[Fig.~\ref{figure.3}(h)], corresponding to the trivial case.

The above results of quench dynamics can be understood from the
tight-binding model given in Eq.~(5), which further takes the form
in the momentum $k$-space
\begin{eqnarray}\label{k}
{H}_k&=& \kappa(a_k^\dagger a_k e^{ikd}e^{i\phi}+a_k^\dagger a_k e^{-ikd}e^{i\phi}+c_k^\dagger c_k e^{ikd}+c_k^\dagger c_k e^{-ikd})\nonumber\\
&& +\eta(a_k^\dagger c_k e^{ikd}+a_k^\dagger c_k
e^{-ikd}e^{i\phi}+c_k^\dagger a_k e^{-ikd}+c_k^\dagger a_k
e^{ikd}e^{i\phi}),
\end{eqnarray}
with $d$ the lattice constant. For $\phi=\pi$, the above
Hamiltonian gives a 1D topological phase known as AIII class
insulator and characterized by a 1D winding
number~\cite{Liu2013,Song2018} (see also details in Appendix B).
The dynamical topological number defined through $\vec
g(\tau_{1,2})$ in quench dynamics precisely corresponds to the 1D
winding number of the above Bloch Hamiltonian. On the other hand,
for $\phi=0$ the above Hamiltonian gives a 1D gapless spin-orbit
coupled band with trivial topology (see Appendix B).

We emphasize the highly nontrivial features of the topological
quench dynamics, which provide the holographic characterization of
the topological phase realized in the ring-resonator system,
namely, the quench dynamics solely in the time-dimension contains
the complete information. The single variable, i.e. the time $t$,
automatically splits into two fundamental time scales, mimicking
the Bloch momenta $\tau$ of the topological band and the residue
time evolution $T$ after quench, respectively, with which the bulk
topology of the system is completely determined \cite{animation}.
Specifically, the pseudo spin dynamics averaged over the time
scale $T$ manifest BIS structure depicted via $\tau$. The
derivative of the $T$-averaged spin dynamics with respect to
$\tau$ across BIS points determines the bulk topology. This result
is in sharp contrast to the conventional characterization of the
non-equilibrium topological invariants, which necessitates the
information in both the time dimension and real momentum space. On
the other hand, this prediction also shows the novelty of
classifying by topological theory the non-equilibrium dynamics,
whose raw features are quite complicated and depend on Hamiltonian
details (Fig.~\ref{figure.2}), but are actually classified by the
underlying universal topological patterns (Fig.~\ref{figure.3})
through the characterization scheme given above~\cite{animation}.

\section{Topological quench dynamics with disorders}

\subsection{Perturbation of disorder in the phase modulator}

\begin{figure}[htbp]
    \centering
    \includegraphics[width=0.98\textwidth ]{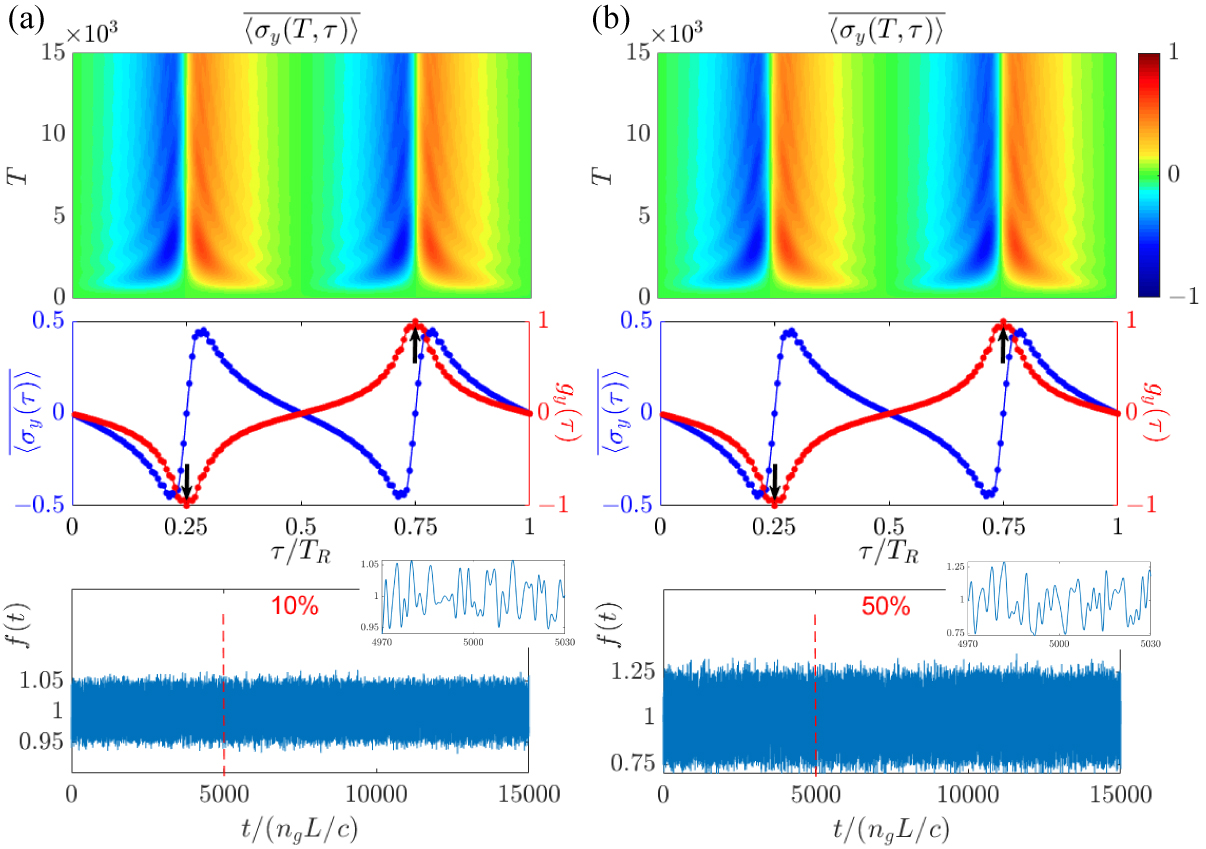}
    \caption{The evolution of averaged spin-polarization $\overline{\langle \sigma_{y}(T,\tau) \rangle}$, the overall spin-polarization $\overline{\langle \sigma_{y}(\tau)\rangle}$, $g_{y}(\tau)$, respectively, with parameters for Fig.~\ref{figure.3}(a) and \ref{figure.3}(b) and disorder function $f(t)$ in $\kappa(t)$ and $\kappa'(t)$ with $\delta=10\%$(a) and $50\%$(b), respectively. Red arrows point to values of $g_y(\tau_{1,2})$.}\label{figure.4}
\end{figure}

In this subsection, we consider the perturbation of topological
quench dynamics from disorders in phase modulators. Such disorders
in the phase modulation can be reflected in modulation strengths,
$\kappa$ and $\kappa'$. We consider that $\kappa$ and $\kappa'$
undergo a random perturbation continuously, which is varying in
time and can be described by $\kappa(t)=\kappa_0\cdot f(t)$,
$\kappa'(t)=\kappa_0'\cdot f(t)$, where $\kappa_0=0.0025\ c/Ln_g$
and $\kappa_0'=0.2\ c/Ln_g$. Here $f(t)=1+\delta r(t)$ is the
disorder function, where $r(t)$ is a time-varying random function
with a range $[-0.5,0.5]$ and $\delta$ represents the disorder
intensity.

Simulations are performed with parameters for
Figs.~\ref{figure.3}(a) and 3(b) and $\delta=10\%$ and $50\%$,
respectively, and results of $\overline{\langle \sigma_{y}(T,\tau)
\rangle}$ together with $\overline{\langle
\sigma_{y}(\tau)\rangle}$ and  $g_{y}(\tau)$ are plotted in
Fig.~\ref{figure.4}, with the corresponding $f(t)$. Compared to
the dynamical pattern in Fig.~\ref{figure.3}(b), evolutions of
$\overline{\langle \sigma_{y}(T,\tau) \rangle}$ with different
disorder $\delta$ show relatively similar profiles. The averaged
spin-polarization pattern and the nontrivial dynamical spin
texture preserve when phase modulators include temporal disorders.
This result can be understand since the temporal disorder in
$\kappa$ and $\kappa'$ does not break the symmetry feature in the
Hamiltonian in Eq.~(\ref{k}), and hence the bulk topology of the
Hamiltonian preserves.

\subsection{Perturbation of disorder in the input source}

\begin{figure}[htbp]
    \centering
    \includegraphics[width=0.98\textwidth ]{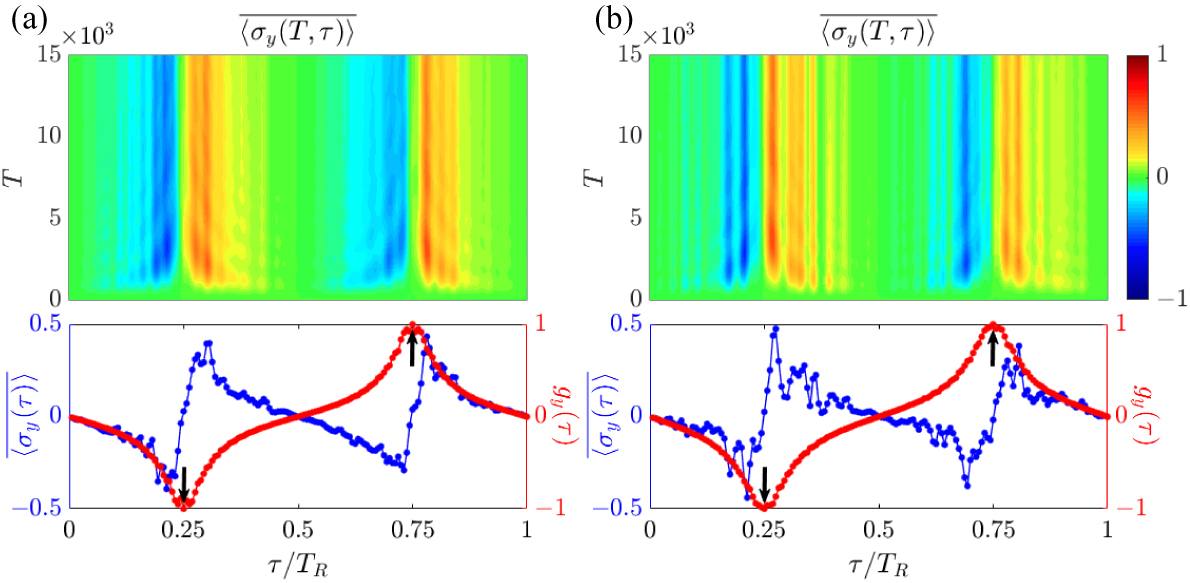}
    \caption{The evolution of averaged spin-polarization $\overline{\langle \sigma_{y}(T,\tau) \rangle}$, the overall spin-polarization $\overline{\langle \sigma_{y}(\tau)\rangle}$, $g_{y}(\tau)$, respectively,  with parameters for Figs.~\ref{figure.3}(a) and \ref{figure.3}(b) and disorder in the input source in Eq.~(\ref{Rand}) with $\delta=5\%$(a) and $10\%$(b), respectively. Red arrows point to values of $g_y(\tau_{1,2})$.}\label{figure.5}
\end{figure}

In simulation for Figs.~\ref{figure.3}(a) and \ref{figure.3}(b),
we prepare the initial state of the system by injecting a
monochromic light at the center frequency $\omega_{C,0}$. In this
subsection, we consider the injected light has disorder in both
intensities and phases for all frequency modes. Such disorder in
the input source can be described by
\begin{equation}\label{Rand}
\left\{
\begin{array}{lr}
E_0^{C,\mathrm{in}}=(1-\delta\cdot R)\cdot s\cdot{\tanh[0.05(t+t_S-t_O/2)]+\tanh[0.05(-t_O/2-t)]}, \\
E_{m\neq 0}^{C,\mathrm{in}}=\delta\cdot R\cdot e^{i2\pi R},\ \ \ \
\ \ \ \ \ \ \ \ 0\leqslant t\leqslant t_s
\end{array}
\right.
\end{equation}
where $\delta$ is the disorder intensity and $R$ gives a random
number in the range $[-0.5,0.5]$.

We perform simulations with same parameters for
Figs.~\ref{figure.3}(a) and \ref{figure.3}(b) and the input source
in Eq.~(\ref{Rand}) with $\delta=5\%$ and $10\%$, respectively.
The corresponding evolutions of $\overline{\langle
\sigma_{y}(T,\tau) \rangle}$ together with $\overline{\langle
\sigma_{y}(\tau)\rangle}$ and  $g_{y}(\tau)$ are plotted in
Fig.~\ref{figure.5}. Although the disorder in the input source
affects more largely the topological quench dynamics for the case
with larger $\delta$, the overall  spin-polarization in
$\overline{\langle \sigma_{y}(\tau)\rangle}$ as well as the
nontrivial dynamical spin texture $g_{y}(\tau)$ still capture the
topological feature of the studied system.

\section{Experimental feasibility}

The quench dynamics is induced by initializing a deep trivial
phase for the topological Hamiltonian, which in principle has a
high experimental feasibility in comparison with the currently
achieved band structure measurement for the resonator ring systems
\cite{r17,Li2020}. In the present study, we do not need to prepare
the initial system to be in the eigenstates of the Hamiltonian,
nor to scan the frequency to match the band energies, which are
however required and were the major challenges for the
conventional band mapping techniques. This essential difference
makes the present quench study be of high feasibility.

Further, the proposed ring resonator system can be achieved in both fiber-based platforms \cite{r11,r17,Li2020} and on-chip lithium niobate photonic designs \cite{c2,c3}, where the parameters in the system can be realized in experiments. In both systems, the conversion efficiency of the electro-optic modulators can reach up to $\sim 2\%$ \cite{c3,c4,c5,c6}, which is sufficient for the proposed system here. The quality factor for the ring is potentially possible at the order of $\sim 10^7-10^8$ with an amplifier for compensation in fiber rings \cite{r11,r17} or the state-of-art integrated  lithium niobate technology \cite{c7}. Therefore, our proposal provides an experimental feasible platform for measuring the quench dynamics and the topological invariants directly from the temporal optical signal in ring resonator, which can lead to significant simplification of performing dynamical characterization of topological quantum phases in different synthetic models. 

\section{Conclusion and discussion}

In summary, we have investigated the topological quench dynamics in a 1D spinful lattice model synthesized in the dimension of frequency of light in ring resonators, and predicted the holographic features of the quench dynamics. 
In particular, we showed that the quench dynamics in time domain
can be generically characterized with two emergent time scales,
with one mimicking the Bloch momentum of the lattice and the other
characterizing the residue time evolution. In this
characterization the quench dynamics, being complicated in the
original time dimension, exhibit universal dynamical topological
patterns which correspond to the bulk topology of the post-quench
Hamiltonian.
The topological quench dynamics is robust against disorders and of
high feasibility in experimental realization. We note that the
approach proposed in this work is generic, and the study can be
readily extended to topological phases in synthetic
high-dimensions, e.g. to the 2D Haldane model~\cite{r28}. In that
case we expect that the multiple fundamental time scales would
emerge in the holographic quench dynamics, with some mimicking the
high-dimensional Bloch momenta and the remaining characterizing
the residue time evolution, for which the complex quench dynamics
can be classified by the exotic dynamical topology and has
profound connection to the bulk topology of the post-quench
Hamiltonian. This work showed a unique way to study the
holographic far-from-equilibrium dynamics, with the rich and
complex topological physics being captured in only the
single-variable, i.e. the time evolution, and shall provided the
insight into the exploration of the high-dimensional topological
phases with quench dynamics in the synthetic photonic crystals.



\section*{Acknowledgements}
This paper was supported by National Natural Science Foundation of
China (11974245, 11825401, and 11761161003), National Key R$\&$D
Program of China (2018YFA0306301 and 2017YFA0303701), Natural
Science Foundation of Shanghai (19ZR1475700), and by the Open
Project of Shenzhen Institute of Quantum Science and Engineering
(Grant No.SIQSE202003). L. Y. acknowledges support from the
Program for Professor of Special Appointment (Eastern Scholar) at
Shanghai Institutions of Higher Learning. X. C. also acknowledges
the support from Shandong Quancheng Scholarship (00242019024).

\begin{appendix}
\appendix

\section{Relation between fields and spin textures}

Signals collected from output waveguide in Fig.~\ref{figure.1}(a)
in the main text are $\psi_A(t)$ and $\psi_C(t)$, respectively.
Spin textures of the one-dimensional pseudo-spin lattice in
Fig.~\ref{figure.1}(c) in the main text include $\langle \sigma_z
\rangle$ and $\langle \sigma_y \rangle$, respectively. The
relation between spin textures $\langle \sigma_z\rangle$, $\langle
\sigma_y\rangle$ and signals $\psi_A$, $\psi_C$ satisfies:
 \begin{eqnarray}\label{p1}
 \langle \sigma_z\rangle&\equiv&
 \begin{pmatrix}
 \psi_{A}^*,\psi_{C}^*e^{-i\Omega t/4}
 \end{pmatrix}
 \begin{pmatrix}
 1 & 0 \\
 0 & -1
 \end{pmatrix}
 \begin{pmatrix}
 \psi_{A}\\
 \psi_{C} e^{i\Omega t/4}
 \end{pmatrix}={\left| {\psi_{A}} \right|^2}-{\left| {\psi_{C}} \right|^2},\\
  \langle \sigma_y\rangle&\equiv&
 \begin{pmatrix}
 \psi_{A}^*,\psi_{C}^* e^{-i\Omega t/4}
 \end{pmatrix}
 \begin{pmatrix}
 0 & -i \\
 i & 0
 \end{pmatrix}
 \begin{pmatrix}
 \psi_{A}\\
 \psi_{C} e^{i\Omega t/4}
 \end{pmatrix}\nonumber\\
 &=& -i\psi_{A}^*\cdot \psi_{C}e^{i\Omega t/4}+i\psi_{C}^*e^{-i\Omega t/4} \cdot \psi_{A}.
 \end{eqnarray}
Here the extra coefficient $e^{\pm i\Omega t/4}$ is from the
frequency offset between rings A and C. It is obvious that Eqs.
(S1) and (S2) can be obtained by subtraction of intensities of two
optical signals and interference between two optical signals,
respectively.

  \section{Comparison with numerical results from tight-binding models}
  \setcounter{figure}{0}
  \renewcommand{\thefigure}{S\arabic{figure}}

  \begin{figure}[htbp]
    \centering
    \includegraphics[width=0.98\textwidth ]{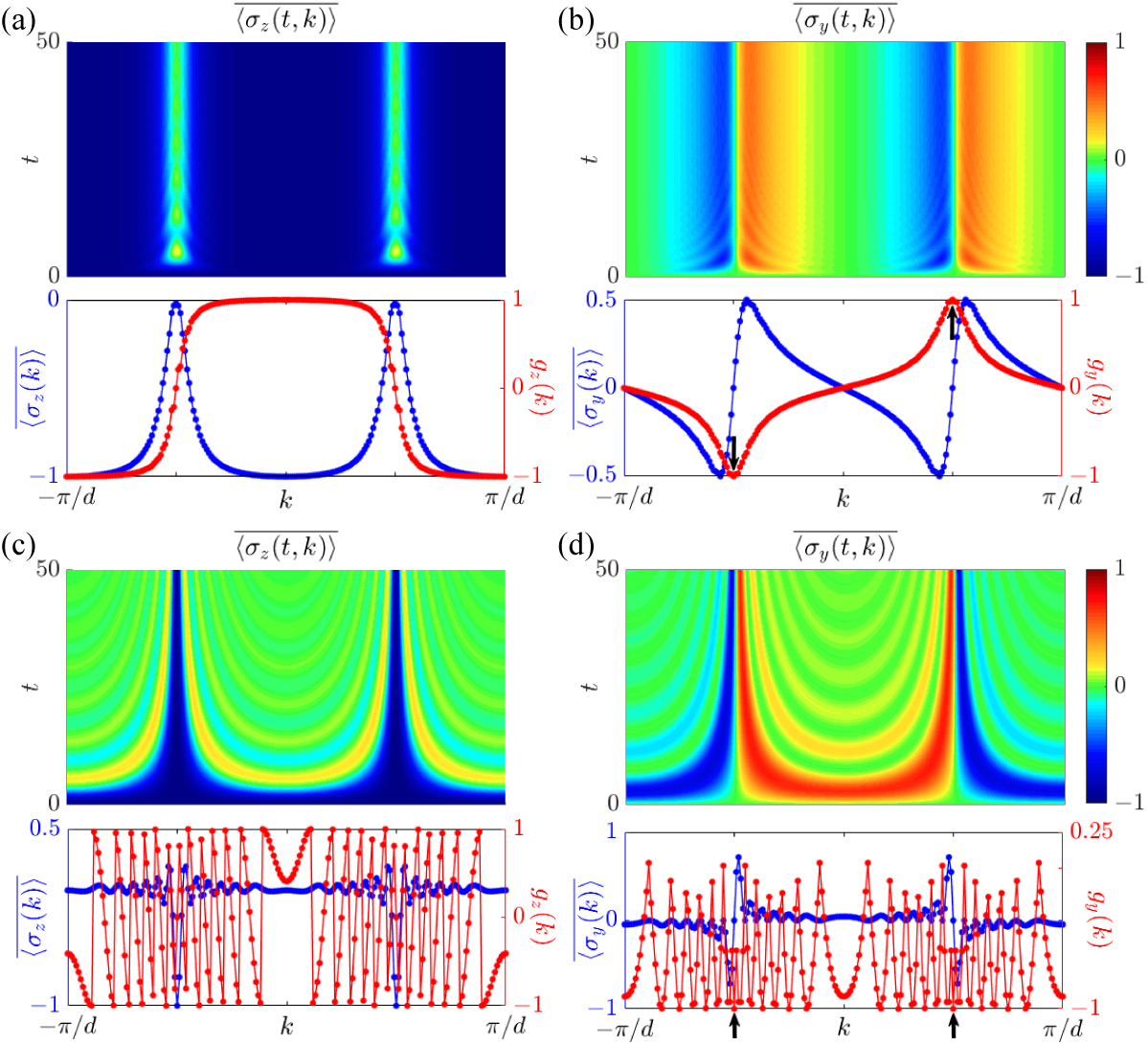}
    \caption{Numerical simulation results of spin textures in the $k$-space. (a) and (b) The evolution of averaged spin-polarization $\overline{\langle \sigma_{z,y}(t,k) \rangle}$, the overall spin-polarization $\overline{\langle \sigma_{z,y}(k)\rangle}$, the dynamical spin texture $g_{z,y}(k)$, respectively, with $\phi=\pi$. (c) and (d) The evolution of averaged spin-polarization $\overline{\langle \sigma_{z,y}(t,k) \rangle}$, the overall spin-polarization $\overline{\langle \sigma_{z,y}(k)\rangle}$, the dynamical spin texture $g_{z,y}(k)$, respectively, with $\phi=0$. Red arrows point to values of $g_y(k=\pm \pi/2d)$.}\label{figure.s1}
  \end{figure}

In this section, we show the numerical simulations of the
corresponding spin textures by numerically solving the
tight-binding model in Eq.~(\ref{r}) in the main text with
$\phi=\pi$ or 0, respectively. The Hamiltonian can be re-written
in the $k$-space:
 \begin{eqnarray}\label{k2}
 {H}_k&=& \kappa(a_k^\dagger a_k e^{ikd}e^{i\phi}+a_k^\dagger a_k e^{-ikd}e^{i\phi}+c_k^\dagger c_k e^{ikd}+c_k^\dagger c_k e^{-ikd})\nonumber\\
 && +\eta(a_k^\dagger c_k e^{ikd}+a_k^\dagger c_k e^{-ikd}e^{i\phi}+c_k^\dagger a_k e^{-ikd}+c_k^\dagger a_k e^{ikd}e^{i\phi}),
 \end{eqnarray}
 where $d$ is the lattice constant. One can rewrite Eq.~(\ref{k2}) in the momentum space $H_k=-2\kappa \cos(kd)\sigma_z-2\eta \sin(kd)\sigma_y$
 when $\phi=\pi$, which gives the 1D AIII class topological insulator with 1D winding number~\cite{Liu2013,Song2018}, and $H_k=2\kappa \cos(kd)I+2\eta\cos(kd)\sigma_x$
 when $\phi=0$. Here $I$ is the unit matrix. The spin textures of the system can then be simulated with initial condition $\langle \sigma_z\rangle=-1, \langle \sigma_y\rangle=0, \langle \sigma_x\rangle=0$, and $\eta=0.2 \kappa$.

 The simulation results of $\langle \sigma_{z,y}(t,k)\rangle$ for $\phi=\pi$ and $0$ are obtained, and corresponding $\overline{\langle \sigma_{z,y}(t,k) \rangle}$, $\overline{\langle \sigma_{z,y}(k)\rangle}$, and $g_{z,y}(k)$ can be defined in the same procedure as that in the main text and then be plotted in Fig.~\ref{figure.s1}. One can notice that these numerical results give the non-trivial feature for $\phi=\pi$ and trivial feature for $\phi=0$, respectively \cite{Zhang2018}. Comparing both results in Fig.~\ref{figure.3} in the main text and Fig.~S1, one finds the consistency in the quench dynamics which gives the same evidences for either non-trivial or trivial case.

 \end{appendix}




\begin{thebibliography}{81}

\bibitem{Hasan2010} M. Z. Hasan and C. L. Kane, Colloquium: Topological insulators, \textit{Rev. Mod. Phys.} {\bf 82}, 3045-3067 (2010).

\bibitem{Qi2011} X.-L. Qi and S.-C. Zhang, Topological insulators and superconductors, \textit{Rev. Mod. Phys.} {\bf 83}, 1057-1110 (2011).

\bibitem{Yan2012} B. Yan and S.-C. Zhang, Topological materials, \textit{Rep. Prog. Phys.} {\bf 75}, 096501 (2012).

\bibitem{Chiu2016} C.-K. Chiu, J. C. Y. Teo, A. P. Schnyder, and S. Ryu, Classification of topological quantum matter with symmetries, \textit{Rev. Mod. Phys.} {\bf 88}, 035005 (2016).

\bibitem{Yan2017} B. Yan and C. Felser, Topological Materials:
Weyl Semimetals, \textit{Annu. Rev. Condens. Matter Phys.} {\bf
8}, 337-354 (2017).

\bibitem{Rudner2013} M. S. Rudner, N. H. Lindner, E. Berg, and M. Levin, Anomalous Edge States and the Bulk-Edge Correspondence for Periodically
Driven Two-Dimensional Systems, \textit{Phys. Rev. X} {\bf 3},
031005 (2013).

\bibitem {Hu2015} W. Hu, J. C. Pillay, K. Wu, M. Pasek, P. Ping Shum, and Y. D. Chong, Anomalous Edge States and the Bulk-Edge Correspondence for Periodically Driven Two-Dimensional Systems, \textit{Phys. Rev. X} {\bf 5}, 011012 (2015).

\bibitem{Mukherjee2017} S. Mukherjee, A. Spracklen, M. Valiente, E. Andersson, P. \"Ohberg, N. Goldman, and R. R. Thomson, Experimental
observation of anomalous topological edge modes in a slowly driven
photonic lattice, \textit{Nat. Commun.} {\bf 8}, 13918 (2017).

\bibitem{Maczewsky2017} L. J. Maczewsky, J. M. Zeuner, S. Nolte, and A. Szameit, Observation of photonic anomalous Floquet topological insulators, \textit{Nat. Commun.} {\bf 8}, 13756 (2017).

\bibitem{Wintersperger2020} K. Wintersperger, C. Braun, F. N. \"Unal, A. Eckardt, M. Di Liberto, N. Goldman, I. Bloch, M. Aidelsburger, Realization of an anomalous Floquet topological system with ultracold atoms, \textit{Nature Phys.} {\bf 16}, 1058-1063 (2020).

\bibitem{VincentLiu2020} H. Hu, B. Huang, E. Zhao, and W. V. Liu, Dynamical Singularities of Floquet Higher-Order Topological Insulators. Phys. Rev. Lett. {\bf 124}, 057001 (2020).

\bibitem{Zhang2020PRL} L. Zhang, L. Zhang, and X.-J. Liu, Unified Theory to Characterize Floquet Topological Phases by Quench Dynamics, \textit{Phys. Rev. Lett.} {\bf 125}, 183001 (2020).

\bibitem{Caio2015} M. D. Caio, N. R. Cooper, and M. J. Bhaseen, Quantum Quenches in Chern Insulators, \textit{Phys. Rev. Lett.} {\bf 115}, 236403 (2015).


\bibitem{Hu2016} Y. Hu, P. Zoller, and J. C. Budich, Dynamical Buildup of a Quantized Hall Response from Nontopological States, \textit{Phys. Rev. Lett.} {\bf 117}, 126803 (2016).

\bibitem{WilsonPRL2016} J.H. Wilson, J.C.W. Song, and G. Rafael, Remnant Geometric Hall Response in a Quantum Quench, Phys. Rev. Lett. {\bf 117}, 235302 (2016).

\bibitem{Wang2017} C. Wang, P. Zhang, X. Chen, J. Yu, and H. Zhai, Scheme to Measure the Topological Number of a Chern Insulator from Quench Dynamics, \textit{Phys. Rev. Lett.} {\bf 118}, 185701 (2017).

\bibitem{Heyl2018} M. Heyl, Dynamical quantum phase transitions: A review, Rep. Prog. Phys. {\bf 81}, 054001 (2018).

\bibitem{Flaschner2018} N. Fl{\"a}schner, D. Vogel, M. Tarnowski, B. S. Rem, D. S. L{\"u}hmann, M. Heyl, J. C. Budich, L. Mathey, K. Sengstock, and C. Weitenberg, Observation of dynamical vortices after quenches in a system with topology, \textit{Nat. Phys.} {\bf 14}, 265-268 (2018).

\bibitem{Song2018} B. Song, L. Zhang, C. He, T. F. J. Poon, E. Hajiyev, S. Zhang, X.-J. Liu, and G.-B. Jo, Observation of symmetry-protected topological band with ultracold fermions, \textit{Sci. Adv.} {\bf 4}, aao4748 (2018).

\bibitem{GongPRL2018} Z. Gong and M. Ueda, Topological Entanglement-Spectrum Crossing in Quench Dynamics, Phys. Rev. Lett. {\bf 121}, 250601 (2018).

\bibitem{McGinley2019} M. McGinley and N. R. Cooper, Classification of topological insulators and superconductors out of equilibrium, \textit{Phys. Rev. B} {\bf 99}, 075148 (2019).



\bibitem{Lu2019} Y.-H. Lu, B.-Z. Wang, and X.-J. Liu, Ideal Weyl semimetal with 3D spin-orbit coupled ultracold quantum gas, \textit{Sci. Bull.} {\bf 65}, 2080-2085 (2020).

\bibitem{QiuX2019} X. Qiu, T.-S. Deng, Y. Hu, P. Xue, and W. Yi, Fixed Points and Dynamic Topological Phenomena
in a Parity-Time-Symmetric Quantum Quench, \textit{iScience} {\bf
20}, 392-401 (2019).

\bibitem{XiePRL2020} D. Xie, T.-S. Deng, T. Xiao, W. Gou, T. Chen, W. Yi, and B. Yan, Topological Quantum Walks in Momentum Space with a Bose-Einstein Condensate, \textit{Phys. Rev. Lett.} {\bf 124}, 050502 (2020).

\bibitem{Hu2020} H. Hu and E. Zhao, Topological Invariants for Quantum Quench Dynamics from Unitary Evolution, \textit{Phys. Rev. Lett.} {\bf 124}, 160402 (2020).

\bibitem{Yu2020} X. Chen, C. Wang, and J. Yu, Linking invariant for the quench dynamics of a two-dimensional two-band Chern insulator, \textit{Phys. Rev. A} {\bf 101}, 032104 (2020).

\bibitem{Slager2020} F. Nur \"{U}nal, A. Bouhon, and R.-J. Slager, Topological Euler Class as a Dynamical Observable in Optical Lattices, \textit{Phys. Rev. Lett.} {\bf 125}, 053601 (2020).

\bibitem{Zhang2018} L. Zhang, L. Zhang, S. Niu, and X.-J. Liu, Dynamical classification of topological quantum phases, \textit{Sci. Bull.} {\bf 63}, 1385-1391 (2018).

\bibitem{Zhang2019} L. Zhang, L. Zhang, and X.-J. Liu, Dynamical detection of topological charges, \textit{Phys. Rev. A} {\bf 99}, 053606 (2019).

\bibitem{Zhang2019b} L. Zhang, L. Zhang, and X.-J. Liu, Characterizing topological phases by quantum quenches: A general theory, \textit{Phys. Rev. A} {\bf 100}, 063624 (2019).

\bibitem{Zhang2019c} L. Zhang, L. Zhang, Y. Hu, N. Sen, and X.-J. Liu, Emergent topology and symmetry-breaking order in correlated quench dynamics, arXiv:1903.09144 (2019).

\bibitem{XLYu2020} X.-L. Yu, L. Zhang, J. Wu, X.-J. Liu, High-order band inversion surfaces in dynamical characterization of topological phases, arXiv:2004.14930 (2020).

\bibitem{Gong2020} L. Li, W. Zhu, and J. Gong, Topological characterization
of higher-order topological insulators with nested band inversion
surfaces, arXiv:2007.05759 (2020).

\bibitem{LiPRA2020} J. Ye and F. Li, Emergent topology under slow nonadiabatic quantum dynamics, Phys. Rev. A {\bf 102}, 042209 (2020).

\bibitem{Sun2018b} W. Sun, C.-R. Yi, B.-Z. Wang, W.-W. Zhang, B. C. Sanders, X.-T. Xu, Z.-Y. Wang, J. Schmiedmayer, Y. Deng, X.-J. Liu, S. Chen, and J.-W. Pan, Uncover Topology by Quantum Quench Dynamics, \textit{Phys. Rev. Lett.} {\bf 121}, 250403 (2018).

\bibitem{Wang2019} Y. Wang, W. Ji, Z. Chai, Y. Guo, M. Wang, X. Ye, P. Yu, L. Zhang, X. Qin, P. Wang, F. Shi, X. Rong, D. Lu, X.-J. Liu, and J. Du, Experimental observation of dynamical bulk-surface correspondence in momentum space for topological phases, \textit{Phys. Rev. A} {\bf 100}, 052328 (2019).

\bibitem{Yi2019} C.-R. Yi, L. Zhang, L. Zhang, R.-H. Jiao, X.-C. Cheng, Z.-Y. Wang, X.-T. Xu, W. Sun, X.-J. Liu, S. Chen, and J.-W. Pan, Observing Topological Charges and Dynamical Bulk-Surface Correspondence with Ultracold Atoms, \textit{Phys. Rev. Lett.} {\bf 123}, 190603 (2019).

\bibitem{Song2019} B. Song, C. He, S. Niu, L. Zhang, Z. Ren, X.-J. Liu, and G.-B. Jo, Observation of nodal-line semimetal with ultracold fermions in an optical lattice, \textit{Nat. Phys.} {\bf 15}, 911-916 (2019).

\bibitem{Ji2020} W. Ji, L. Zhang, M. Wang, L. Zhang, Y. Guo, Z. Chai, X. Rong, F. Shi, X.-J. Liu, Y. Wang, and J. Du, Quantum Simulation for Three-Dimensional Chiral Topological Insulator, \textit{Phys. Rev. Lett.} {\bf 125}, 020504 (2020).

\bibitem{Xin2020} T. Xin, Y. Li, Y.-a. Fan, X. Zhu, Y. Zhang, X. Nie, J. Li, Q. Liu, and D. Lu, Quantum Phases of Three-Dimensional Chiral Topological Insulators on a Spin Quantum Simulator, \textit{Phys. Rev. Lett.} {\bf 125}, 090502 (2020).

\bibitem{Niu2020} J. Niu, T. Yan, Y. Zhou, Z. Tao, X. Li, W. Liu, L. Zhang, S. Liu, Z. Yan, Y. Chen, and D. Yu, Simulation of Higher-Order Topological Phases and Related Topological Phase Transitions in a Superconducting Qubit, arXiv:2001.03933 (2020).

\bibitem{BChen2021} B. Chen, S. Li, X. Hou, F. Ge, F. Zhou, P. Qian, F. Mei, S. Jia, N. Xu, and H. Shen, Digital quantum simulation of Floquet topological phases with a solid-state quantum simulator, Photonics Research {\bf 9}, 81-87 (2021).

\bibitem{r1} D. I. Tsomokos, S. Ashhab, and F. Nori, Using superconducting qubit circuits to engineer exotic lattice systems, \textit{Phys. Rev. A} {\bf 82}, 052311 (2010).

\bibitem{r2} O. Boada, A. Celi, J. I. Latorre, and M. Lewenstein, Quantum Simulation of an Extra Dimension, \textit{Phys. Rev. Lett.} {\bf 108}, 133001 (2012).

\bibitem{r3} D. Juki$\mathrm{\acute{c}}$ and H. Buljan, Four-dimensional photonic lattices and discrete tesseract solitons, \textit{Phys. Rev. A} {\bf 87}, 013814 (2013).

\bibitem{r4} L. Yuan, Q. Lin, M. Xiao, and S. Fan, Synthetic dimension in photonics, \textit{Optica} {\bf 5}, 1936 (2018).

\bibitem{r5} T. Ozawa, H. M. Price, A. Amo, N. Goldman et al, Topological photonics, \textit{Rev. Mod. Phys.} {\bf 91}, 015006 (2019).

\bibitem{YuanNanophotonics} D. Leykam, and L. Yuan, Topological phases in ring resonators: recent progress and future prospects, \textit{Nanophotonics} {\bf 9}, 4473 (2020).

\bibitem{r6} X. -W. Luo, X. Zhou, C. -F. Li, J. -S. Xu et al, Quantum simulation of 2D topological physics in a 1D array of optical cavities, \textit{Nat. Commun.} {\bf 6}, 7704 (2015).

\bibitem{r7} L. Yuan, Y. Shi, and S. Fan, Photonic gauge potential in a system with a synthetic frequency dimension, \textit{Opt. Lett.} {\bf 41}, 741-744 (2016).

\bibitem{r8} T. Ozawa, H. M. Price, N. Goldman, O. Zilberberg, and I. Carusotto, Synthetic dimensions in integrated photonics: From optical isolation to four-dimensional quantum Hall physics, \textit{Phys. Rev. A} {\bf 93}, 043827 (2016).

\bibitem{r9} L. Yuan, Q. Lin, A. Zhang, M. Xiao, X. Chen, and S. Fan, Photonic Gauge Potential in One Cavity with Synthetic Frequency and Orbital Angular Momentum Dimensions, \textit{Phys. Rev. Lett.} {\bf 122}, 083903(2019).

\bibitem{syn-atom-1} A. Celi, P. Massignan, J. Ruseckas, N. Goldman, I. B. Spielman, G. Juzeli$\mathrm{\bar{u}}$nas, and M. Lewenstein, Synthetic Gauge Fields in Synthetic Dimensions, \textit{Phys. Rev. Lett.} {\bf 112}, 043001 (2014).

\bibitem{r10} E. Lustig, S. Weimann, Y. Plotnik, Y. Lumer, M. A. Bandres, A. Szameit, and M. Segev, Photonic Topological Insulator in Synthetic Dimensions, \textit{Nature} {\bf 576}, 356-360 (2019).

\bibitem{r11} A. Dutt, Q. Lin, L. Yuan, M. Minkov, M. Xiao, and S. Fan, A single photonic cavity with two independent physical synthetic dimensions, \textit{Science} {\bf 367}, 59-64 (2020).

\bibitem{syn-atom-2} M. Mancini, G. Pagano, G. Cappellini, L. Livi, M. Rider, J. Catani, C. Sias, P. Zoller, M. Inguscio, M. Dalmonte, and L. Fallani, Observation of chiral edge states with neutral fermions in synthetic Hall ribbons, \textit{Science} {\bf 349},
1510-1513 (2015).

\bibitem{syn-atom-3} B. K. Stuhl, H.-I. Lu, L. M. Aycock, D. Genkina, and I. B. Spielman, Visualizing edge states with an atomic Bose gas in the quantum Hall regime, \textit{Science} {\bf 349}, 1514-1518 (2015).

\bibitem{r12} Q. Lin, M. Xiao, L. Yuan, and S. Fan, Photonic Weyl point in a two-dimensional resonator lattice with a synthetic frequency dimension, \textit{Nat. Commun.} {\bf 7}, 13731 (2016).

\bibitem{r13} Q. Lin, X. -Q. Sun, M. Xiao, S. -C. Zhang, and S. Fan, A three-dimensional photonic topological insulator using a two-dimensional ring resonator lattice with a synthetic frequency dimension, \textit{Sci. Adv.} {\bf 4}, eaat2774 (2018).

\bibitem{r14} B. A. Bell, K. Wang, A. S. Solntsev, D. N. Neshev, A. A. Sukhorukov, and B. J. Eggleton, Spectral photonic lattices with complex long-range coupling, \textit{Optica} {\bf 4}, 1433-1436 (2017).

\bibitem{r15} C. Qin, F. Zhou, Y. Peng, D. Sounas, X. Zhu, B. Wang, J. Dong, X. Zang, A. Al$\mathrm{\grave{u}}$, and P. Lu, Spectrum Control through Discrete Frequency Diffraction in the Presence of Photonic Gauge Potentials, \textit{Phys. Rev. Lett.} {\bf 120}, 133901 (2018).

\bibitem{r16} L. J. Maczewsky, K. Wang, A. A. Dovgiy, A. E. Miroshnichenko, A. Moroz, M. Ehrhardt, M. Heinrich, D. N. Christodoulides, A. Szameit, and A. A. Sukhorukov, Synthesizing multi-dimensional excitation dynamics and localization transition in one-dimensional lattices, \textit{Nat. Photonics} {\bf 14}, 76-81 (2020).

\bibitem{r17} A. Dutt, M. Minkov, Q. Lin, L. Yuan, D. A. B. Miller, and S. Fan, Experimental band structure spectroscopy along a synthetic dimension, \textit{Nat. Commun.} {\bf 10}, 3122 (2019).

\bibitem{Li2020} G. Li, Y. Zheng, A. Dutt, D. Yu, Q. Shan, S. Liu, L. Yuan, S. Fan, and X. Chen, Dynamic band structure measurement in the synthetic space, \textit{Sci. Adv.} {\bf 7} eabe4335 (2021).

\bibitem{s1} L. Yuan, and S. Fan, Bloch oscillation and unidirectional translation of frequency in a dynamically modulated ring resonator, \textit{Optica} {\bf 3}, 1014-1018 (2016)

\bibitem{s2} L. Yuan, Q. Lin, M. Xiao, A. Dutt, and S. Fan, Pulse shortening in an actively mode-locked laser with parity-time symmetry, \textit{APL Photonics} {\bf 3}, 086103 (2018).

\bibitem{s3} D. Yu, L. Yuan, and X. Chen, Isolated Photonic Flatband with the Effective Magnetic Flux in A Synthetic Space including the Frequency Dimension, \textit{Laser Photonics Rev.} {\bf 14}, 2000041 (2020).

\bibitem{s4} Z. Yang, E. Lustig, G. Harari, Y. Plotnik, Y. Lumer, M. A. Bandres, and M. Segev, Mode-Locked Topological Insulator Laser Utilizing Synthetic Dimensions, \textit{Phys. Rev. X} {\bf 10}, 011059 (2020).

\bibitem{r28} L. Yuan, M. Xiao, Q. Lin, and S. Fan, Synthetic space with arbitrary dimensions in a few rings undergoing dynamic modulation, \textit{Phys. Rev. B} {\bf 97}, 104105 (2018).

\bibitem{a2} A. Yariv and P. yeh, Photonics: Optical Electronics in Modern Communications (Oxford University, New York, 2007).


\bibitem{b1} H. A. Haus, Waves and Fields in Optoelectronics (Prentice-Hall, Inc., Englewood Cliffs, NJ, 1984).

\bibitem{b2} B. E. A. Saleh and M. C. Teich, Fundamentals of Photonics (Wiley-Interscience, Hoboken, NJ, 2007).

\bibitem{Liu2013} X.-J. Liu, Z.-X. Liu, and M. Cheng, Manipulating Topological Edge Spins in a One-Dimensional Optical Lattice, \textit{Phys. Rev. Lett.} {\bf 110}, 076401 (2013).

\bibitem{animation} See Supplemental Material at http://link.aps.org/supplemental/..., which includes a cartoon illustrating how the bulk topology of the system can be completely determined from the information of quench dynamics solely in the time dimension.

\bibitem{c2} M. Zhang, C. Wang, Y. Hu, A. Shams-Ansari, T. Ren, S. Fan, M. Lon$\mathrm{\check{c}}$ar, Electronically programmable photonic molecule, \textit{Nat. Photonics} {\bf 13}, 36-40 (2019).

\bibitem{c3} C. Wang, M. Zhang, X. Chen, M. Bertrand, A. Shams-Ansari, S. Chandrasekhar, P. Winzer, and M. Lon$\mathrm{\check{c}}$ar, Integrated lithium niobate electro-optic modulators operating at CMOS-compatible voltages, \textit{Nature} {\bf 562}, 101-104 (2018).

\bibitem{c4} L. D. Tzuang, K. Fang, P. Nussenzveig, S. Fan, and M. Lipson, Non-reciprocal phase shift induced by an effective magnetic flux for light, \textit{Nat. Photonics} {\bf 8}, 701-705 (2014).

\bibitem{c5} C. Wang, M. Zhang, B. Stern, M. Lipson, and M. Lon$\mathrm{\check{c}}$ar, Nanophotonic lithium niobate electro-optic modulators, \textit{Opt. Express} {\bf 26}, 1547-1555 (2018).

\bibitem{c6} C. Reimer, Y. Hu, A. Shams-Ansari, M. Zhang, and M. Lon$\mathrm{\check{c}}$ar, High-dimensional frequency crystals and quantum walks in electro-optic microcombs, arXiv:1909.01303 (2019).

\bibitem{c7} B. Desiatov, A. Shams-Ansari, M. Zhang, C. Wang, and M. Lon$\mathrm{\check{c}}$ar, Ultra-low-loss integrated visible photonics using thin-film lithium niobate, \textit{Optica} {\bf 6}, 380-384 (2019).

\end{thebibliography}

\end{document}